\documentclass{pasa}%

\usepackage{graphicx}

\title[Non-inertial effects in disk galaxies]{The Effects of Inertial Forces on the Dynamics of Disk Galaxies}

\author[Gomel and Zimmerman]{R. Gomel$^1$, T. Zimmerman$^2$
\affil{$^1$School of Physics and Astronomy, Tel Aviv University, Israel, roygomel6@gmail.coml}%
\affil{$^2$School of Physics and Astronomy, Tel Aviv University, Israel, stomerzi@gmail.com}
}%

\usepackage{aas_macros}
\usepackage{hyperref} 
\hypersetup{colorlinks,citecolor=blue,linkcolor=blue,urlcolor=blue}

\begin{document}

\begin{frontmatter}
\maketitle

\begin{abstract}
When dealing with galactic dynamics, or more specifically, with galactic rotation curves, one basic assumption is always taken: the frame of reference relative to which the rotational velocities are given is assumed to be inertial. In other words, fictitious forces are assumed to vanish relative to the observational frame
of a given galaxy. It might be interesting, however, to explore the outcomes of dropping that assumption; that is, to search for signatures of non-inertial behavior in the observed data.
In this work, we show that the very discrepancy in galaxy rotation curves could be attributed to non-inertial effects. We derive a model for spiral galaxies that takes into account the possible influence of fictitious forces and find that the additional terms in the new model, due to fictitious forces, closely resemble dark halo profiles. Following this result, we apply the new model to a wide sample of galaxies, spanning a large range of luminosities and radii.
It turns out that the new model accurately reproduces the structures of the rotation curves and provides very good fittings to the data.
\end{abstract}

\end{frontmatter}

\section{Introduction}

\begin{figure}[t]
	\includegraphics[width=\columnwidth]{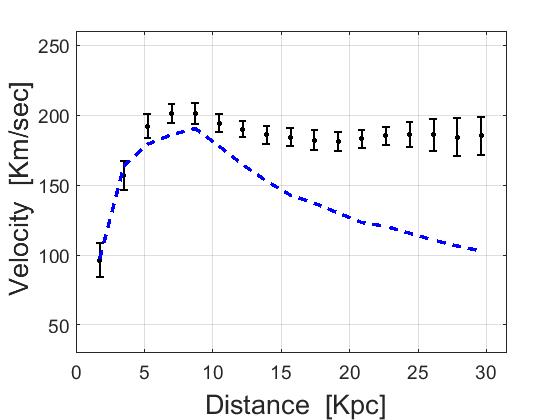}
    \caption{A measured rotation curve (black points with error-bars) and its Newtonian prediction (dashed blue line).
		The predicted curve is based on visible matter alone. 
		The inconsistency with the observed data is known as the discrepancy in galaxy rotation curves. 
		The data (NGC 4157) was taken from McGaugh's Data Pages \citep{McGaugh2017}.}
    \label{fig:RC_example}
\end{figure}

One of the major tools to analyze the dynamics and mass distributions of disk galaxies is a Rotation Curve \cite{SofueRubin2001}.
It presents the variation in the orbital velocities within a galaxy at different distances from the center. Its introduction, however, gave rise to a fundamental discrepancy: Newtonian predictions seemed to be incompatible with actual observations. According to Newtonian dynamics, the rotational velocities at the outskirts of galaxies
should decrease with distance. However, actual observations did not show such a trend \citep{Babcock1939,Salpeter1978, Rubin1980, SancisiAlbada1987}. An illustration of the discrepancy is presented in Figure~\ref{fig:RC_example}.\\

% Characteristic Rotation Curve figure

Two different approaches to dealing with the discrepancy have emerged over the years. One approach poses that Newtonian models are missing additional mass (i.e., dark matter), while the second approach argues that the very application of Newton's laws is invalid in these cases \citep{Milgrom1983}. The vast majority of astrophysicists support the first explanation as dark matter explains much more than only rotation curves. Dark matter is a major ingredient in current cosmology, explaining the CMB \citep{Jarosik2011}, structure formation \citep{DelPopolo2007}, the discrepancies in galaxy clusters \citep{Massey2010}, merging galaxy clusters \citep{Markevitch2004}, and more.\\

In the field of galaxy dynamics, and specifically in disk galaxies with measured rotation curves, a basic assumption has always been made: the frame of reference, relative to which the observed rotational velocities are given, is assumed to be inertial (e.g., the black points in Figure~\ref{fig:RC_example} are assumed to be given relative to an inertial frame of reference). Note that the rotational velocities, by definition, must be given relative to some frame of reference. Since the frame is assumed to be inertial, the influence of fictitious forces should not be taken into account in the models. Indeed, the blue curve in Figure~\ref{fig:RC_example} (i.e., the baryonic rotation curve) does not include any contribution of fictitious forces. It is derived from the gravitational attraction of the visible matter alone.\\

This current work explores the consequences of dropping the basic assumption. What if the observed rotational velocities (of each and every galaxy) cannot be treated as inertial? It should be noted that the assumption regarding the observed velocities has always been taken implicitly. It was not discussed in early works in this field. In the next section we show that by explicitly defining the frame, relative to which the observed rotational velocities are given, the possibility of the frame being non-inertial is revealed.\\

Motivated by this insight, we derive a new model for the rotational velocities. The new model includes an additional term due to the possible presence of fictitious forces. It turns out that the contribution of the additional term to the rotational velocities
is similar to the contribution of common dark halo profiles (e.g., NFW, Burkert). This surprising result might have implications other than those directly related to rotation curves. However, these implications should be discussed separately and independently. This work focuses only on presenting the result and confirming its  universality. It does not deal with the possible implications on other fields of research.\\

The paper is organized as follows: Section \ref{Section2} derives a new model for the rotational velocities, i.e., a model for disk galaxies that takes into account the influence of inertial forces. Section \ref{Section3} demonstrates that this model is effectively similar to common dark halo profiles.
Section \ref{Section4} uses this model to fit a wide sample of rotation curves,  and in Section \ref{Section5} we conclude and discuss future directions.

\section{Deriving a New Model}\label{Section2}

One of the main results of the general theory of relativity (GR) is the insight that a gravitational field could arise not only by means of mass distributions but also due to the state of motion of the frame itself \citep{Einstein1920}. That is, with respect to an ``accelerating'' or a ``rotating'' frame of reference, a gravitational field would appear. Although the weak-field approximation is used in the regime of galactic dynamics, this basic insight still holds:
a gravitational field would appear if the frame is non-inertial. The
$\it frame$, in this context, is simply the observational frame of reference, relative to which the data is presented. The $\textit{gravitational field}$, in this context, is simply the field due to classical fictitious forces. If indeed such a field exists relative to the observational frame of a given galaxy, it may ``play the role'' of a dark halo. However, can the observational frame be non-inertial?\\

When dealing with a terrestrial phenomena (e.g., analyzing the motion of a basketball), it is quite clear how to define the local inertial frame. We a priori know that Newtons laws would hold if we are to describe the dynamics relative to a frame of reference rigidly attached to the ground. In exactly the same manner, we a priori know that fictitious forces must be added to the equations if we wish to describe the motion of the ball relative to a rotating carousel. However, when analyzing the rotational velocities of a given galaxy one cannot, in general, assume that the observed velocities are a priori being given relative to a local inertial frame.\\

\begin{figure}[t]
	\includegraphics[width=\columnwidth]{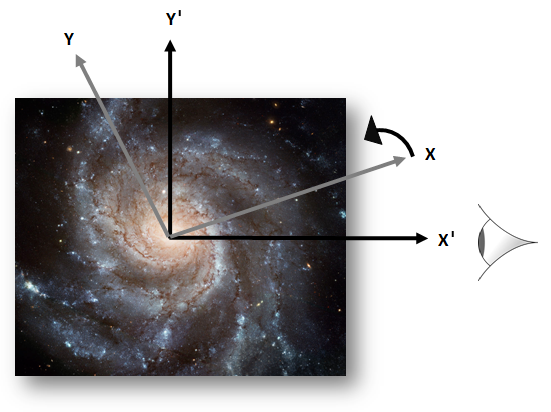}
    \caption{The measured velocities presented in RCs are valid only with respect to $K'$, by definition. If $K'$ is not inertial, then the relative motion between $K$ (the inertial frame) and $K'$ (the observational frame) includes only rotation.
		For the sake of simplicity, in this figure the observer is located on the galactic plane. The definitions of $K$ and $K'$, however, hold for any observer.}
    \label{fig:K_Kp}
\end{figure}

Let us denote the system of coordinates relative to which the rotational data of a given disk galaxy is presented (i.e., the ``observational'' frame) as $K'$. In Appendix \ref{AppendixA}, we show that this frame of reference can be chosen such that its $x'$--$y'$ plane is aligned with the galactic plane and its origin with the galactic center (see Figure~\ref{fig:K_Kp}). We also show that $K'$ is constrained: the line-of-sight connecting the observer and the specific galaxy must be ``fixed'' relative to that frame. This is a direct outcome of using line-of-sight velocities when deriving a rotation curve.\\

Next, let us introduce frame $K$, which is assumed to be inertial, at least locally. That is, a frame of reference relative to which the behavior of bodies in a specific disk galaxy could be described by applying Newton's laws of motion. Without loss of generality, this frame can always be arranged with its $x$--$y$ plane aligned with the galactic plane and its origin aligned with the galactic center
(see Figure~\ref{fig:K_Kp}). The reason for this stems from the
nature of the central-force problem: if Newton's laws are to be valid
relative to that frame, then, assuming a central-force problem, each
body would preserve its plane of motion. Therefore, the whole galactic plane would be ``stationary'' relative to that frame. It is now only a matter of definition to call it ``the $x$--$y$ plane''.\\

% System of Co-ordinates figure

Without any further information, one has no basis to assume the coincidence of $K$ and $K'$. There is a possible relative motion between the two systems: an angular velocity of one system relative to the $z$-axis of the other. Therefore, the observational frame $K'$ could in principle be non-inertial.\\

In order to model the rotational velocities relative to $K'$, we take the following approach: we treat the possible angular velocity between $K$ and $K'$ (of a given galaxy) as a free parameter. We use this degree of freedom to find a relation between the velocities $v_{K'}(r)$ and $v_{K}(r)$. By using such a relation and the known model for $v_{K}(r)$, the velocities $v_{K'}(r)$ could be extracted. The model for $v_{K}(r)$, in this context, is simply the Newtonian model using the visible mass distribution (e.g., the blue curve in Figure~\ref{fig:RC_example}).\\

Finding such a relation is quite straightforward. However, it would be beneficial to derive it from the relation between the accelerations since the effects of the fictitious terms would explicitly be presented. In addition, note that a classical transformation is justified since the weak field approximation is used in the regime of galactic dynamics and the characteristic velocities are much smaller than the speed of light. Given a frame of reference $K'$, rotating at a constant angular velocity $\boldsymbol{ \omega}$  relative to an inertial frame $K$, the relation between the accelerations in the two frames is given by:
\begin{equation}
	\mathbf{a_{K'}} = \mathbf{a_{K}}- 2\boldsymbol{\omega}\times\mathbf{v_{K'}} - \boldsymbol{\omega} \times (\boldsymbol{\omega} \times \mathbf{r_{K'}}) \label{eq:fict},
\end{equation}
where $\mathbf{a_{K}}$ ($\mathbf{a_{K'}}$) is the acceleration of a body relative to $K$ ($K'$), $\boldsymbol{\omega}$ is the constant angular velocity at which system $K'$ revolves relative to system $K$, $\mathbf{v_{K'}}$  is the velocity of the body relative to $K'$ 
and $\mathbf{r_{K'}}$ is the position of the body relative to $K'$. 
Note that the Coriolis term,  $-2\boldsymbol{\omega}\times\mathbf{v_{K'}}$, 
and the centrifugal term, $-\boldsymbol{\omega} \times (\boldsymbol{\omega} \times \mathbf{r_{K'}})$,
are present in the transformation.\\

In our case it would be more convenient to define a positive angular velocity in the opposite direction, as can be seen in Figure~\ref{fig:K_Kp}.
Therefore, taking $\omega$ $\rightarrow$ $-\omega$ and assuming that a body performs a counterclockwise uniform circular motion, leads to 
the following relation in the $\boldsymbol{\hat{r}}$ direction:
\begin{equation}
	(-\frac{v_{K'}^2}{r}) \boldsymbol{\hat{r}} = (-\frac{v_{K}^2}{r}- 2\omega v_{K'} + \omega^2 r) \boldsymbol{\hat{r}} \label{eq:fict_vel},
\end{equation}
where $v_{K}$  ($v_{K'}$) is the magnitude of the rotational velocity relative to $K$ ($K'$) ,  $\omega$ is the constant angular velocity at which system $K$ revolves around the $z'$-axis of system $K'$ and $r$ is the radius of the circular motion.\\

At this stage, a useful sanity check could already be done. We recall that a dark halo produces an extra force in the inward direction.
Therefore, we have to make sure that the two fictitious forces combined, could also act in the inward direction. Looking at \mbox{Equation~(\ref{eq:fict_vel}),} one can see that this condition
is satisfied when $\omega^2 r < 2\omega v_{K'}$, or $ 0 < \omega < \frac{2v_{K'}}{r}$. For a given galaxy and a given observed rotation curve $v_{K'}(r)$, this inequality sets limits on $\omega$.
Within these limits, one can find an $\omega$ which produces an inward net force in $K'$ at any radius. Whether this additional force could imitate the behavior of dark halo profiles is the subject of the next section.\\

Next, solving Equation~(\ref{eq:fict_vel}), we find:
\begin{equation}
	v_{K'}(r)=v_{K}(r)+\omega r\label{eq:V_tr}.
\end{equation}
Equation~(\ref{eq:V_tr}) is the essence of this work. The velocities $v_{K'}(r)$ are the new model for the rotational velocities. 
The idea is that the expected Keplerian behavior (for the rotational velocities) could be measured only relative to $K$. The observational frame $K'$, however, differs from $K$ in the general case (i.e., when $\omega \ne 0$). In that case, the model for the rotational velocities should include an additional linear term ($\omega r$).
In the next section we show that the additional term can effectively ``play the role'' of a dark halo. In Section \ref{Section4} we show that the new model can be used to fit a large sample of rotation curves. Let us mention again that the model developed here is classical. If one wishes to arrive at a more general relation, compatible with GR for any radius, then a derivation based on Einstein field equations should be used.

\section{Inertial Forces as an Effective Dark Halo}\label{Section3}

% Effective DM
\begin{figure*}
	\begin{center}
		\includegraphics[scale=0.27]{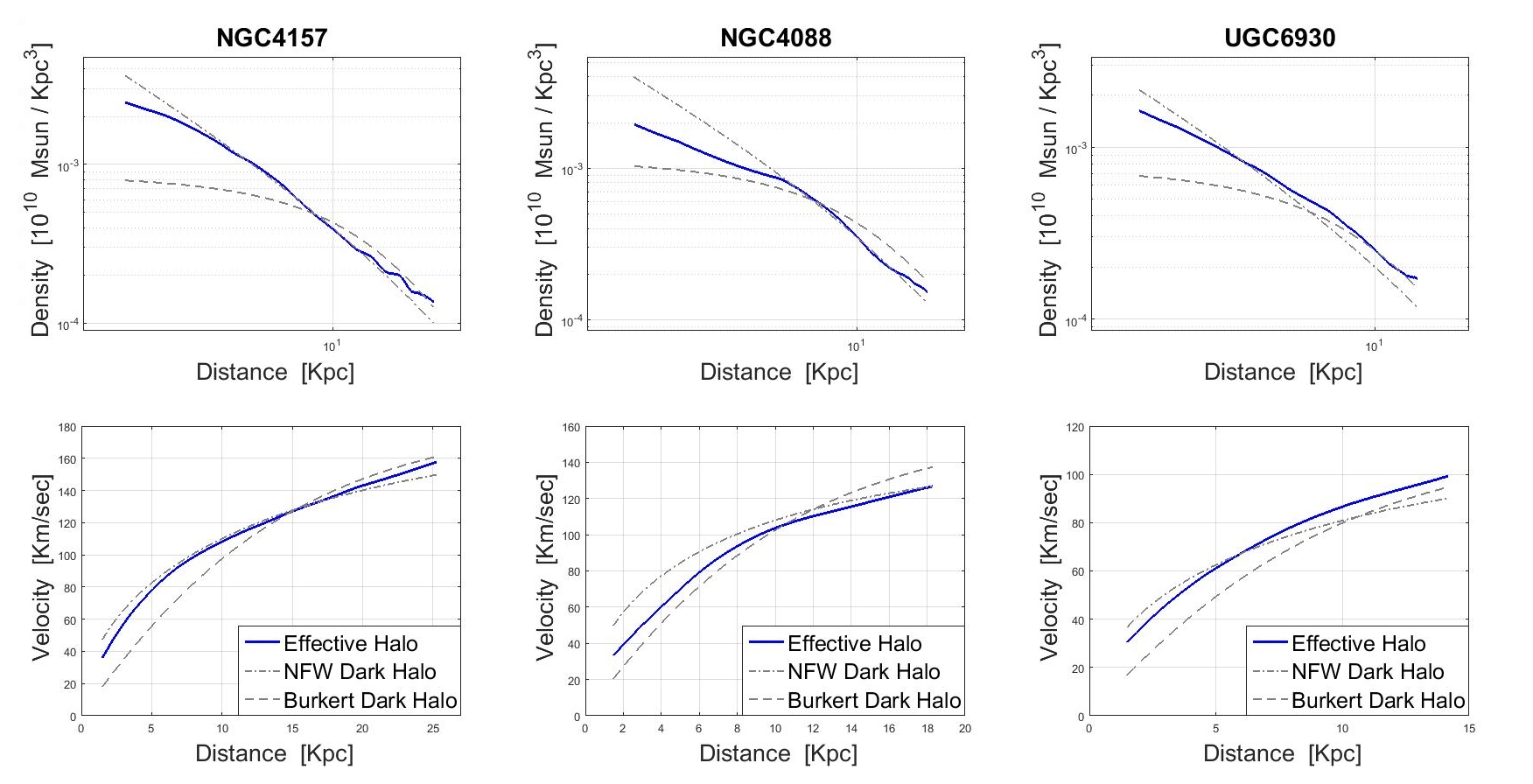}
		\caption{Density distributions of Burkert,  NFW, and the effective halo are plotted in the upper panels. 
The corresponding rotational velocities are given in the lower panels.
 The Burkert and the NFW halos use two free parameters: 
The virial mass, Mvir [$10^{10}$ Msun] and the concentration parameter, c. 
The effective halo has one parameter, namely $\omega$ [$10^{-16}$ rad/sec]. 
Left Panels: NGC 4157 (NFW: Mvir = 170 , c = 5;   Burkert: Mvir = 120 , c = 12;   effective halo: $\omega$ = 1.12). 
Middle Panels: NGC 4088 (NFW: Mvir = 70 , c = 7;   Burkert: Mvir = 70 , c = 13.5;   effective halo: $\omega$ = 1.04). 
Right Panels: UGC 6930 (NFW: Mvir = 40 , c = 5;   Burkert: Mvir = 30 , c = 11.5;   effective halo: $\omega$ = 1.17).}
		\label{fig:Eff_DM}
	\end{center}
\end{figure*}

It is well known that dark halos are used successfully to fit all sorts of RCs \citep{Carignan1985, Corbelli2000, DeBlok2008}. It would be beneficial, therefore, to explore whether our model can imitate 
the behavior of some common dark-halo profiles. In this section, we wish to derive an artificial dark component, with a velocity distribution $V_{dark}(r)$, which would be equivalent to the new model. To that end, the following condition must be satisfied:
\begin{equation}
	\sqrt{V_{baryonic}^{2}(r) + V_{dark}^{2}(r)} = V_{inertial}(r)+\omega r\label{eq:equality}.
\end{equation}
The left-hand side is simply the model for the rotational velocities in the presence of a dark halo. The right-hand side is our new model for the rotational velocities. Using an artificial dark component with a distribution $V_{dark}(r)$ which satisfies the above equation is effectively the same as using the new model.\\

Before proceeding, note that the terms $V_{baryonic}(r)$ and $V_{inertial}(r)$ refer to exactly the same thing---the circular velocities produced by the baryonic component as seen from an inertial frame of reference (e.g., the blue curve in Figure~\ref{fig:RC_example}). Taking this into account and
extracting $V_{dark}(r)$ from Equation~(\ref{eq:equality}) gives:
\begin{equation}
	V_{dark}(r) = \sqrt{\omega^2 r^2 + 2\omega r V_{baryonic}(r)}\label{Vdark},
\end{equation}
where $V_{dark}(r)$ denotes the circular velocities of bodies orbiting an artificial dark halo; that is, a dark halo which produces the same velocity field (and gravitational field) as fictitious forces do.\\

It would be beneficial to derive the artificial-halo's density distribution as well. A short derivation which assumes a spherically symmetrical distribution (i.e., like Burkert, NFW) gives:
\begin{equation}
	\rho_{dark}(r) = \frac{\omega}{4\pi G}(3\omega + 4\frac{V_{baryonic}(r)}{r} + 2V'_{baryonic}(r))\label{Rho_dark}.
\end{equation}\\

As can be seen in Equations (\ref{Vdark}) and (\ref{Rho_dark}), the effective halo is characterized by some unique features.
First, the distribution of baryons is embedded within the effective halo. This coupling between dark and baryonic matter in spiral galaxies has been discussed extensively in the literature \citep{PSS1996, SB2000, Swaters2012, McGaugh2014}. Here, this connection is a natural characteristic of the effective halo.
Second, the effective halo includes a single free parameter (i.e., $\omega$). Most of the common dark-halo profiles include two free parameters (e.g., the virial mass and the concentration parameter).
However, these two parameters are known to be observationally \mbox{correlated  \citep{Burkert1995, Jimenez2003}.} Therefore, in practice, common dark halos as well as effective dark halos can be expressed with a single free-parameter. It is worth mentioning that the mass-to-light parameter (hiding in $V_{baryonic}(r)$) was not counted as an additional parameter since it has to be set also when real dark halos are used.\\

Next, we plot the effective-halo profile together with common dark-halo profiles. The dependence of the effective halo on $V_{baryonic}(r)$, however, compels us to do so only per given galaxy. In Figure~\ref{fig:Eff_DM} we use three different galaxies in order to compare between a Burkert dark halo \citep{Burkert1995},
an NFW dark halo \citep{NFW1997} and our effective halo. For each galaxy we draw the density distribution and the circular-velocity distribution of each halo. In order to set the free parameters of each halo we use the RCs data of \cite{McGaugh2017}. For each galaxy, we choose the halo parameters which produce the best-fit to the data, using a fixed M/L for the baryonic component. The process of fitting the data is explained in Section \ref{Section4}. The values and the actual distributions are presented in the figure.\\

Looking at each panel of Figure~\ref{fig:Eff_DM}, the main result is the similarity between the distributions. In general, it seems that non-inertial effects can reproduce the behavior of dark halos. 
Considering the density profiles, one can notice that the  effective-halo distribution in the inner regions is less steep than that shown by NFW. In the central region, the effective halo mostly follows the NFW behavior, while in the outer regions it has a ``tail''. Let us note that the effective halo cannot reproduce all the different shapes that NFW or Burkert support. It can imitate their shapes only when their parameters are tuned to fit real data.
Specifically, in the cases we present here, the parameters' values are consistent with a ``maximal'' disk, i.e., the contribution of the dark halos in the inner regions is small.\\

\section{Applying the New Model on a Sample of Disk Galaxies}\label{Section4}

In Section \ref{Section2}, we suggested that the new model for the rotational velocities should include an additional term ($\omega r$).
This additional term arises in non-inertial systems. It represents the additional tangential velocity emerging in these systems. 
In Section \ref{Section3}, we demonstrated that such a term can effectively ``play the role'' of a dark halo. Its contribution to the circular velocities (at least in the examined test cases) is practically similar to the contribution of NFW or Burkert. The next natural step would be to fit a large number of rotation curves using the new model.\\

For this purpose we use the publicly available database of \cite{McGaugh2017}. The database includes galaxies with extended 21-cm rotation curves spanning a large range of luminosities and radii. Besides the observed rotation curves, the Newtonian velocity components of the stellar disk $v_{disk}(r)$, the galactic bulge $v_{bulge}(r)$, and the gaseous disk $v_{gas}(r)$ of each galaxy are also available in the database. Those stand for the expected circular velocities produced by each galactic component. For the sake of simplicity, only galaxies without a dedicated bulge component were selected in this work.\\

% RCs - numerically
\begin{figure*}
	\begin{center}
		\includegraphics[scale=0.165]{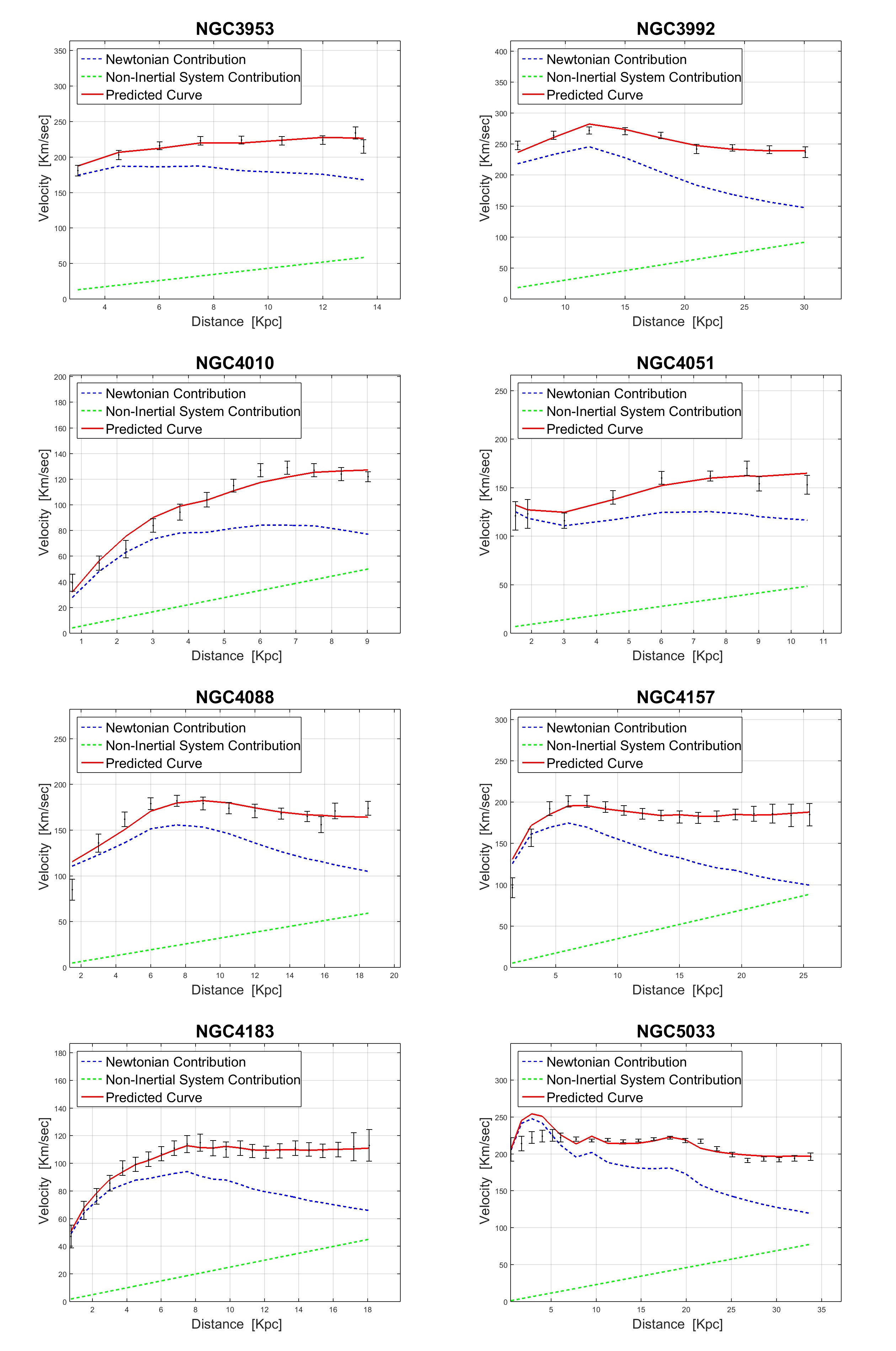}
		\caption{RCs for the different galaxies are presented. 
In each panel one can see the measured values (black error bars) together with the predicted curve (red line). 
The dashed blue curve corresponds to the Newtonian term $v_{k}(r)$ while the dashed green line corresponds to the linear correction term $\omega r$.}
		\label{fig:8_RCs-main}
	\end{center}
\end{figure*}

Let us now explicitly introduce the new model and discuss its different terms. In accordance with Equation~(\ref{eq:V_tr}), the predicted rotation curve of a given galaxy is given by:
\begin{equation}
V_{predicted}(r)= \sqrt{(M/L)\cdot v_{disk}^{2}(r)+v_{gas}^{2}(r)}+\omega r\label{eq:fitting_model},
\end{equation}
where $M/L$ is the stellar-disk mass-to-light ratio, $v_{disk}(r)$ is the normalized stellar-disk contribution, $v_{gas}(r)$ is the gaseous-disk contribution, and $\omega$ is the angular velocity between the frames, as was discussed in previous sections. The model contains, therefore, two free parameters: $M/L$ and $\omega$. The contribution of the detectable matter to the rotation curve is encapsulated within the first term whereas the contribution of the frame being non-inertial is given by $\omega r$.\\

The publicly-available velocity distribution $v_{disk}(r)$ of each galaxy was numerically derived from the corresponding light distribution of that galaxy. The light distribution of a galaxy is assumed to trace the stellar mass distribution. Given the mass distribution (of the stellar disk), the gravitational and the velocity field of the disk component could be derived. The ratio between the mass and the light distributions is therefore required. Being highly uncertain, it is represented by a free parameter, namely $M/L$. The publicly-available velocity distribution $v_{g}(r)$ of each galaxy was numerically derived from the observed neutral hydrogen. The gaseous disk component can make a dominant contribution, especially in low-luminosity galaxies. The disk and the gas components of each galaxy, $v_{disk}(r)$ and $v_{gas}(r)$ are available online in \cite{McGaugh2017} as numerical data files. More details on the extraction of these components can be found in \cite{Sanders2002}.\\

In order to fit the new model (Equation (\ref{eq:fitting_model})) to a given observed rotation curve, we use the least-squares method. We search for the values of $M/L$ and $\omega$ that minimize $\chi^2$ in each galaxy; i.e., the best-fit values of the two parameters. A table that summarizes the best-fit parameters of each galaxy can be found in Appendix \ref{AppendixB}. A fair selection of our fittings to rotation curves is presented in Figure~\ref{fig:8_RCs-main}. The rest of the fittings could also be found in Appendix \ref{AppendixB}.\\

Looking at Figure~\ref{fig:8_RCs-main}, the first outcome is the 
dominance of the Newtonian curve in the inner regions of the RCs.
Inside the inner regions, where the linear term is still small, 
the Newtonian term almost fits the data. In fact, it is similar to the ``maximum-disk'' approach, where the dark-halo contribution in the inner regions is still small and the baryonic curve almost fits the data. This result is very important. When using a dark halo without adopting the maximal-disk approach, a degeneracy between the disk and the halo is usually \mbox{revealed \citep{Dutton2005, Geehan_et_al2006}.}
Our model eliminates the degeneracy but still provides very accurate results.\\

Another natural outcome is the consistency of the fits with Renzo's experimental \mbox{law \citep{Sancisi2004}.} This law states that every time
a feature arises in the radial light distribution, the rotation curve shows a corresponding feature. These features should clearly be evident in the inertial (baryonic) velocities $V_K(r)$, as those were directly derived from the light distributions. In our case, the predicted curves are obtained simply by adding a linear term to the inertial velocities; thus the features are preserved. NGC 1560 (see Appendix \ref{AppendixB}) is a good example: it can clearly be seen that the small ``bump'' in the data (at 5 Kpc) has a corresponding ``bump'' in the predicted curve. Common dark halos usually struggle to predict these ``bumps'' \citep{Gentile2010}.\\

The next point deals with the $M/L$ values. 
Looking at Table 1 (given in \mbox{Appendix \ref{AppendixB}}) one can notice that the fitted values are quite reasonable. The median value of
the population is 1.5 {[}$M_{sun}/L_{sun}${]} while the median absolute deviation is 0.5 {[}$M_{sun}/L_{sun}${]}.\\

% Correlations Figure
\begin{figure*}
	\centering
	\includegraphics[scale=0.15]{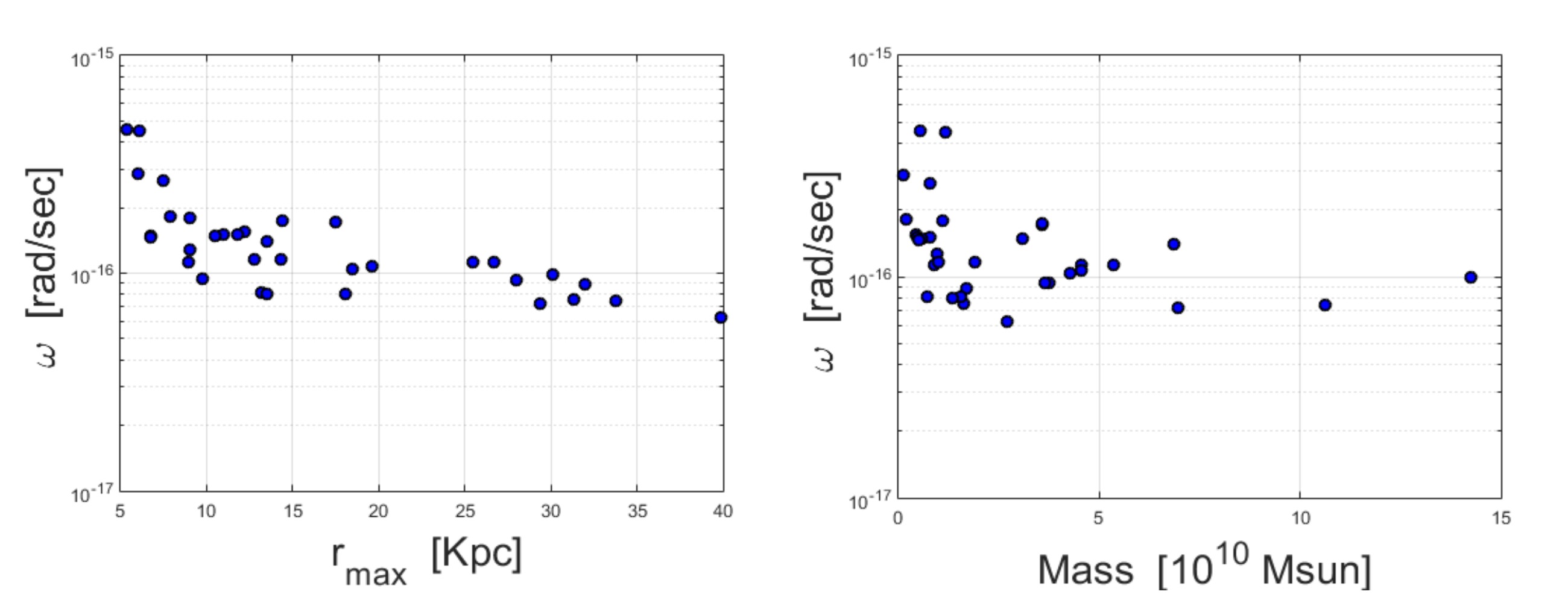}
	\caption{Left panel: a scatter plot of the best-fitted $\omega$'s vs $r_{max}$ (the last point of the RC).
                         Right Panel: a scatter plot of $\omega$ vs the galaxy's mass. 
                         Both plots were produced using the sample of galaxies summarized in Appendix B.}
	\label{fig:Corr}
\end{figure*}

Another point is the $\chi^2$ values. In most of the galaxies, the values seem to be reasonable and to indicate that the fittings are accurate. However, the photometry-based mass models are known to be limited (e.g., sensitive the extraction of the surface photometry,
to the bulge-disk decomposition, and to irregular morphologies). It is clear that very detailed observed rotation curves (e.g., as in the case of M33) could not be well fitted by these models in terms of $\chi^2$.\\

The last point deals with the new parameter, $\omega$. 
This value represents the amount in which the local inertial system of a given galaxy is rotating relative to the observational system of that galaxy. It is clear that only positive $\omega$'s are viable in order to fit the data. This means that the local inertial frame $K$ must revolve relative to the observational frame $K'$ in the same direction as the direction of the revolving matter in that galaxy, i.e., in the direction of the galactic spin. This observation should be examined and analyzed in future works, but let us present a preliminary possible mechanism for this effect.\\

The recent observations of \cite{Lee2019} used the line-of-sight velocity measurements of 434~disk galaxies to show that the rotational direction of a galaxy is coherent with the average motion of its neighbors. Using our notations, this means that the neighbors of a given disk galaxy were found to orbit (on average) around the origin of frame $K'$. The ``neighbors'', in this context, are galaxies located up to 15 Mpc from the given disk galaxy. Such a rotation could, in general, be responsible for the introduction of inertial forces in $K'$ \citep{Thirring1918, Brill_Cohen1966}, or, in different words, be responsible for the ``dragging'' of the local inertial frame $K$ in the direction of the galactic spin. However, let us clarify: explaining or modeling the surprising results found by \cite{Lee2019} is not included in the scope of this work. Those are very preliminary. We mention their results as a basis for future works. Note that the dragging effect suggested here is caused by the background and not by the galaxy itself.\\

It would also be beneficial to check whether the values of $\omega$ correlate with other galactic quantities. In Figure~\ref{fig:Corr}, we present scatter plots of $\omega$ against two galactic parameters: the mass and the radius. In the left panel one can see that the larger the radius of a galaxy (represented here by the last point of the RC), the smaller the value of $\omega$. In the right panel one can see a similar trend (although weaker) with the galaxy's total (baryonic) mass.

\section{Summary and Future Directions}\label{Section5}

Thus far, two main approaches have been taken in order to deal with the discrepancy in galaxy rotation curves: changing the underlying laws of physics or adding more mass to the detectable mass distribution. The additional inward attraction, in both cases, result in corrected rotational velocities. In the scope of this work, we have shown that the different dynamics in the presence of inertial forces also lead to corrected rotational velocities. In fact, these new rotational velocities strongly resemble those obtained by using dark-halo profiles.\\

The motivation to proposing that fictitious forces may arise relative to the observational frame of a galaxy originates in one basic argument: the observational frame ($K'$) could not be treated with certainty as inertial from the first place. This assumption was not tested and was taken as ``ground truth'' mostly due to its implicit nature. Testing such an assumption may require a model that takes into account the various possible phenomena (in different scales) that affect the determination of a local inertial frame.\\

In this work we have demonstrated that a single degree of freedom is sufficient in order to represent the possible relative motion between the observational frame ($K'$) and the local inertial frame ($K$) of a given disk galaxy. Relying on this degree of freedom, we have demonstrated that the additional gravitational field in $K'$, due to fictitious forces, closely resembles the gravitational field of a dark halo. Additionally, a model that predicts the (non-inertial) rotational velocities in $K'$ was developed. Applying the new model to a wide sample of RCs produced very accurate results.\\

The idea that the observed velocities might be non-inertial
can be generalized, in principle, to other types of galaxies.
However, the specific model we developed here is applicable only to
disk galaxies. It might be interesting, therefore, to search for equivalent models in the field of elliptical galaxies: that is, to explore the discrepancy in measured velocity dispersions. Another interesting direction for further investigation is to analyze galactic stability in the presence of inertial forces. If those could indeed imitate the gravitational behavior of dark halos then their role in N-body simulations and other measures of galactic stability could be important.\\

The discrepancy in galaxy rotation curves is probably one of the most surprising discrepancies in astrophysics. It seems to puzzle us even today, decades after it was first discovered.

\section*{acknowledgements}
We wish to thank the anonymous reviewers whose comments and suggestions helped to improve and clarify this paper. We wish to thank Stacy McGaugh for sharing the rotation-curves data and for valuable advice. Additionally, we would like to thank Meir Zeilig-Hess and Oran Ayalon for helpful insights.\\

\section*{Data availability}
The data underlying this article are available in DATA SECTION, at http://astroweb.case.edu/ssm/data.
It includes extended 21-cm rotation curves as well as B-band photometry for the sample of galaxies analyzed in this work.

% \bibliographystyle{pasa-mnras}
% \bibliography{1r_lamboo_notes}

\begin{appendix}
\newpage
\section{Defining the Observational Frame }\label{AppendixA}

The goal of this appendix is to remind the reader how the rotational velocities of a galaxy
are derived (i.e., the data points in an RC), and to provide a well-defined system of coordinates
relative to which these rotational velocities are given.\\

In general, the process of deriving a rotation curve could be separated into two steps: 
first, measure the line-of-sight velocities in a galaxy (i.e., the 2D velocity field);
second, convert these line-of-sight velocities into rotational velocities \citep{Swaters2009}.
Before proceeding, a crucial clarification is needed:
the line-of-sight velocity of an object is a component of the object's total velocity along the line of sight.
The object's total velocity, by definition, must be given relative to some frame of reference.
Therefore, if an object's total velocity is to be calculated from the line-of-sight velocity (e.g., by some geometrical considerations),
then it is necessary to provide the corresponding frame as well.\\

The most basic conversion between line-of-sight velocities and circular velocities
uses only the line-of-sight velocities located on the main axis of the galaxy's projected image
and assumes a single fixed value for the disk inclination \citep{DeBlok2008}. 
In this case, the conversion is simply given by:
\begin{equation}
v_{c}(r)=\frac{v(r)-v_{sys}}{sin(i)}\label{Appendix A Eq},
\end{equation}
where $v_{c}(r)$ are the rotational velocities (e.g., the black data
points in Figure~\ref{fig:RC_example}), $v(r)$
are the line-of-sight velocities, $v_{sys}$ is the galaxy's systemic
velocity, and $i$ is the inclination angle of the galactic plane
($i=90^{o}$ for edge-on galaxies).
Relying on this simple transformation would not change 
the conclusions we draw at the end of this section.\\

\begin{figure}[h]
	\centering
	\includegraphics[width=\hsize]{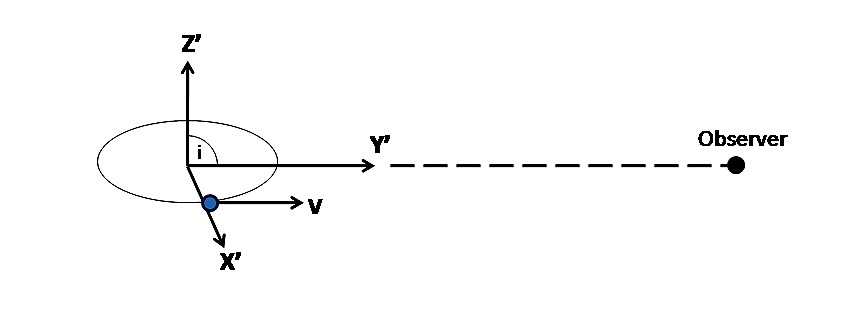}
	\caption{
                         A stationary distant observer is located on the $y'-axis$ (i.e. $v_{sys}=0$, $sin(i)=1$).
                         Such an observer would measure a value of $V$ for the line of sight velocity,
                         thus, relying on relation \ref{Appendix A Eq}, would extract a value of $V$ 
                         for the rotational velocity. As can be seen in the figure, this value is valid
                         relative to the system $K'$. At any other frame which is rotating relative to $K'$, the rotational
                         velocity would take a different value.          
                         }
	\label{fig:velocities_transformation}
\end{figure}

As noted before, the reference frame 
(relative to which the rotational velocities $v_{c}(r)$ are given)
is in fact determined when using such a relation.
Actually, it is not fully determined but strongly constrained. 
It turns out that any reference frame in which the systemic velocity of a galaxy vanishes
and the line-of-sight is stationary (i.e., does not rotate),
can be regarded as valid.
To illustrate this, let us focus on the (unrealistic) example
in which $v_{sys}=0$, $sin(i)=1$.
In this case, relation (\ref{Appendix A Eq}) states that the rotational velocities
are equal to the measured line-of-sight velocities.
However, as can be seen in Figure~\ref{fig:velocities_transformation},
this statement holds only relative to system $K'$,
where the line of sight is stationary.
In any reference frame which is rotating relative to $K'$,
the rotational velocity would take a different value, the line of sight would not be fixed
and relation (\ref{Appendix A Eq}) would not hold.
The reader may take a minute to be fully convinced at this point.\\

Now, let us define one specific reference frame which 
satisfies the above requirement (i.e., a frame, relative to which the rotational velocities are given). We define a frame whose
fundamental plane coincides with the galaxy plane 
and its origin coincides with the galactic center.
We also set the frame in such a way that the line-of-sight of the observer is stationary
(in the simple case of an edge-on galaxy, the primary axis is pointing towards the observer).
This frame of reference is denoted in the main text as $K'$.
It turns out that frame $K'$ might be non-inertial (see Section \ref{Section2}).

% Diagram

\section{RC fittings}\label{AppendixB}
The RC fittings for the entire sample as well as the best-fitted parameters of each galaxy are given in the next pages.\\

% A: RCs - numerically
\begin{figure*}
	\begin{center}
		\includegraphics[scale=0.12]{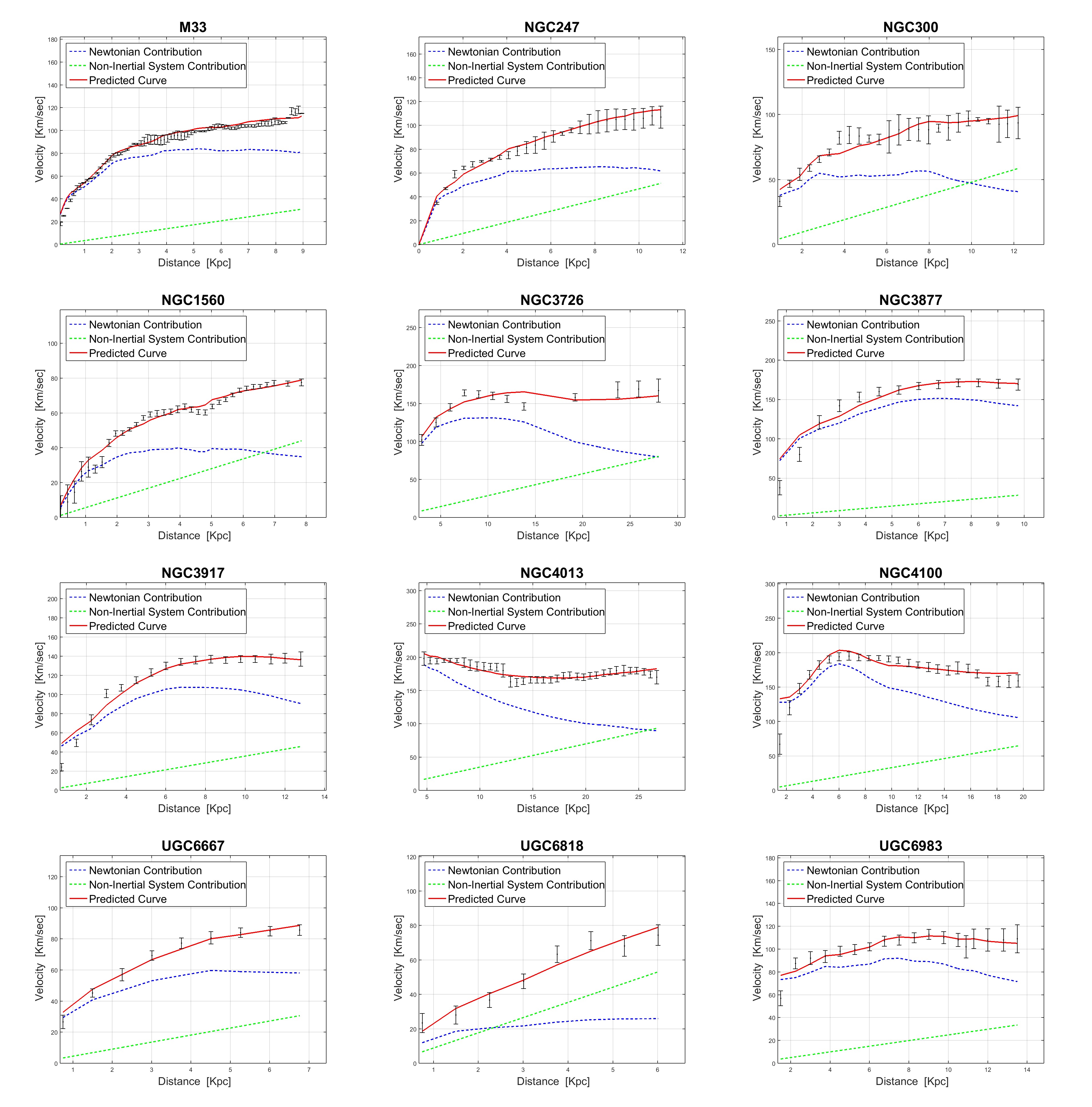}
	\end{center}
\end{figure*}

% B: RCs - numerically
\begin{figure*}
	\begin{center}
		\includegraphics[scale=0.12]{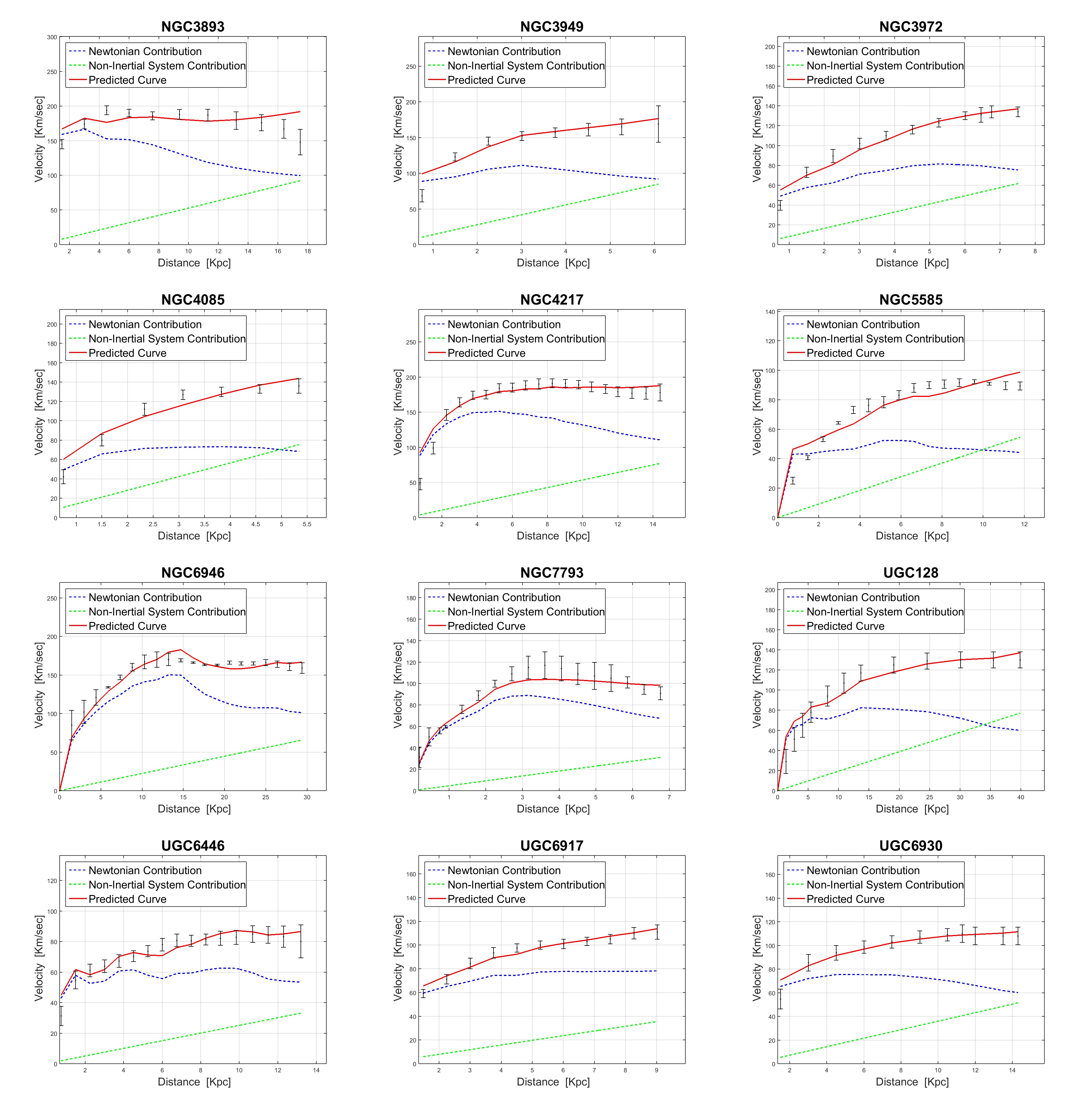}
		\caption{RCs for the different galaxies are presented. 
In each panel one can see the measured values (black error bars) together with the predicted curve (red line). 
The dashed blue curve corresponds to the Newtonian term $v_{k}(r)$ while the dashed green line corresponds to the linear correction term $\omega r$.}
	\end{center}
\end{figure*}

\begin{table}[h]
\begin{center}
	\begin{tabular}{@{}cccccccc@{}}
		\hline
		Galaxy & $\omega$$^a$ & $M/L$$^b$ & $\chi^2$$^c$ &
		Galaxy & $\omega$ & $M/L$ & $\chi^2$\\
		\hline
		M33           & 1.12  & 0.97 & 118    & NGC 4085  & 4.58  & 0.50 & 3.19\\
		NGC 1003  & 0.75  & 0.54 & 5.56  & NGC 4088  & 1.04  & 1.16 & 1.48\\ 
		NGC 1560  & 1.82  & 2.57 & 2.38  & NGC 4100  & 1.07  & 2.36 & 2.33\\
		NGC 247    & 1.51  & 1.76 & 3.74  & NGC 4157  & 1.12  & 2.27 & 0.83\\
		NGC 300    & 1.55  & 0.96 & 1.14  & NGC 4183  & 0.81  & 1.45 & 0.1\\ 
		NGC 3726  & 0.93  & 1.19 & 3.42  & NGC 4217  & 1.73  & 1.72 & 2.99\\
		NGC 3769  & 0.89  & 1.77 & 4.69  & NGC 5033  & 0.74  & 5.06 & 4.44\\ 
		NGC 3877  & 0.94  & 1.78 & 3.01  & NGC 5585  & 1.50  & 0.87 & 14.6\\ 
		NGC 3893  & 1.71  & 1.43 & 3.53  & NGC 6946  & 0.72  & 0.79 & 8.76\\ 
		NGC 3917  & 1.16  & 1.58 & 4.38  & NGC 7793  & 1.49  & 1.40 & 1.22\\ 
		NGC 3949  & 4.51  & 0.48 & 2.78  & UGC 128    & 0.63  & 3.49 & 1.05\\ 
		NGC 3953  & 1.40  & 2.25 & 0.69  & UGC 6446  & 0.81  & 1.70 & 1\\ 
		NGC 3972  & 2.66  & 1.00 & 1.67  & UGC 6667  & 1.47  & 1.74 & 0.83\\ 
		NGC 3992  & 0.99  & 4.26 & 1.13  & UGC 6818  & 2.86  & 0.16 & 1.03\\ 
		NGC 4010  & 1.80  & 1.37 & 1.38  & UGC 6917  & 1.28  & 2.01 & 0.94\\ 
		NGC 4013  & 1.13  & 2.90 & 0.95  & UGC 6930  & 1.17  & 1.30 & 0.52\\ 
		NGC 4051  & 1.49  & 1.11 & 1.02  & UGC 6983  & 0.80  & 3.11 & 0.87\\
		\hline
	\end{tabular}

\caption{Best-fitted parameters of each galaxy. $^a$  $\omega$ is given in  $[10^{-16} rad/sec]$. $^b$  $M/Ls$ are given in the B-band $[M_{sun}/L_{sun}]$. $^c$  These are the reduced $\chi^2$ values of each fit.  }

	\end{center}
\end{table}

\end{appendix}

\end{document}